\documentclass[aps,prl,twocolumn,superscriptaddress,amsmath,amssymb]{revtex4}
\usepackage{graphicx}
\usepackage{color}
\begin{document}
\title{Effect of intense, ultrashort laser pulses on DNA plasmids in their native state: strand breakages induced by {\it in-situ} electrons and radicals}
\author{J. S. D$'$Souza}
\affiliation{UM-DAE Centre for Excellence in Basic Sciences, University of Mumbai, Kalina Campus, Mumbai 400 098, India}
\author{J. A. Dharmadhikari}
\affiliation{Tata Institute of Fundamental Research, 1 Homi Bhabha Road, Mumbai 400 005, India}
\author{A. K. Dharmadhikari} 
\affiliation{Tata Institute of Fundamental Research, 1 Homi Bhabha Road, Mumbai 400 005, India}
\author{B. J. Rao}
\affiliation{Tata Institute of Fundamental Research, 1 Homi Bhabha Road, Mumbai 400 005, India}
\author{D. Mathur}
\email{atmol1@tifr.res.in}
\affiliation{Tata Institute of Fundamental Research, 1 Homi Bhabha Road, Mumbai 400 005, India}

\begin{abstract}
Single strand breaks are induced in DNA plasmids, pBR322 and pUC19, in aqueous media exposed to strong fields generated using ultrashort laser pulses (820 nm wavelength, 45 fs pulse duration, 1 kHz repetition rate) at intensities of 1-12 TW cm$^{-2}$. The strong fields generate, {\it in situ}, electrons and radicals that induce transformation of supercoiled DNA into relaxed DNA, the extent of which is quantified. Introduction of electron and radical scavengers inhibits DNA damage; results indicate that OH radicals are the primary (but not sole) cause of DNA damage. 
\end{abstract}
\maketitle

Studies of the interaction of radiation with biological matter have long focused on quantification of energy transfer from the radiation field into irradiated matter. The physics governing primary energy deposition has been understood for some decades \cite{inokuti} and finds applications in the biomedical sciences. Information now exists that readily enables not only deduction of macroscopic patient doses in radiotherapy \cite{Horton} but also microdosimetric doses within single cells \cite{rossi}. However, while it is now routinely possible to quantify the energy that is deposited in a given volume of irradiated matter, there remains a gap in knowledge as to the subsequent sequence of events that unfold. As a result quantitative insights into how a given dose of radiation induces biological effects continue to be elusive. Since the most important radiation damage is that caused to the genome, it is clear that the focus of experimental studies must be on DNA. The breaking of single and double DNA strands constitutes potentially the most lethal damage at the cellular level. For long it has been thought that such damage is caused by exposure of living matter to high-energy radiation \cite{radiation_damage} that ionizes the sugar-phosphate backbone. However, about a decade ago, Sanche and coworkers pioneered a new line of condensed-phase experiments that offered indications that even electrons possessing only a few eV's of energy might induce single strand breaks (SSB) and double strand breaks (DSB) \cite{sanche_early} through the formation of temporary negative ion states that subsequently dissociate.  

Breakage of DNA strands by low energy electrons is of interest as such electrons are copiously produced along tracks of ionizing radiation, typically about 10$^4$ electrons per MeV that is deposited \cite{radiation}. Li {\it et al.} \cite{Li} carried out model calculations in which a section of DNA backbone was represented by two deoxyribose (sugar) rings that were connected by a phosphate; {\it ab initio} computations of adiabatic potential energy surfaces of the neutral and the anion revealed that bond rupture is thermodynamically favorable for activation barriers of $\sim$10 kcal mol$^{-1}$. In solution phase, the energetics are likely to be different because of solvation effects. Even though a solvated electron reacts slowly with dialkyl phosphate anions, computations have shown that direct damage to the DNA backbone by low energy electrons may be so fast that it actually precedes solvation \cite{Li}. Indeed, an electron with only $\sim$5 eV energy would lead to formation of DNA multiple transient anion states which decay into damaged structures involving SSB and DSB \cite{sanche_early,Huels}. In earlier experiments on water,ionization with ultraviolet photons was the precursor for pre-hydrated electrons which rapidly solvate on timescales $<$1 ps \cite{earlier_work}; due to solvation, the radius of such electrons is reduced tenfold from the $\sim$30 {\AA} value for a pre-hydrated electron. The work that we report is in the liquid phase and it follows a qualitatively different strategy wherein a strong optical field is made to undergo space-time focusing, resulting in self-steepening and plasma formation such that electrons and radicals are generated {\it in situ} via multiphoton, tunneling and dissociation processes. 

We have explored electron-and radical- mediated damage to DNA in its native, aqueous state. Such damage manifests itself in the creation of relaxed forms of DNA which we monitor using gel electrophoresis. We observe that upon addition of electron scavengers like 5-bromo-uracil and melatonin, there is a significant reduction in the population of relaxed species. Similar reduction is obtained upon addition of mannitol and sodium acetate, scavengers of OH radicals \cite{PNAS}. Damage is, therefore, essentially caused by electrons and radicals that are produced as the aqueous water+DNA is exposed to strong optical fields. The electrons are produced {\it in situ} by ultrashort (45 fs duration) pulses of 820 nm light of incident intensities ($I$) in the range 1-12 TW cm$^{-2}$ at a repetition rate of 1 kHz. It is known that self-focusing of incident light leading to filamentation may increase localized intensities to levels beyond 12 TW cm$^{-2}$. On the basis of recent time-resolved shadowgraphy experiments and solutions of nonlinear Schr\"odinger equations \cite{Mindardi}, it is estimated (S. Minardi and A. Couairon - private communication) that localized electron densities of $>$10$^{19}$ cm$^{-3}$ are obtained within the 1 cm interaction region in our experiments.  Descriptions of our apparatus can be found in recent reports on supercontinuum generation in crystals and macromolecular media \cite{whitelight}. The intensity of light that is incident on the DNA+water sample is high enough for us to invoke the optical Kerr effect wherein the total refractive index ($n$) comprises a linear and an intensity-dependent nonlinear portion, $n = n_o+n_2I$. The laser beam's Gaussian intensity profile then maps to a refractive index profile $n = n_o+n_2I~exp(-2r^2/w_o^2)\approx n_o+n_2I(1- 2r^2/w_o^2)$. The radial dependence of the phase of the propagating beam results in self-focusing within the irradiated aqueous medium until high enough intensity is attained for multiphoton ionization (MPI) to occur. MPI-generated electrons, in turn, contribute to de-focusing such that propagation through the medium proceeds in a series of self-focusing-de-focusing events, giving rise to filamentation \cite{PUILS,filamentation} (for a cogent review, see \cite{physrep}). In our laser intensity regime, multiple filaments are formed \cite{PUILS} within the irradiated DNA-containing liquid (Fig. 1a) and the interaction length is 1 cm (interaction volume is $\sim$4$\times$10$^{-5}$ cm$^{3}$). The plasmid DNA (pUC19 and pBR322) used in our work are from a commercial source (Merck-Millipore). These plasmids were suspended in 2 $\ell$ of de-ionized water in dialysis bags with a 12 kDa molecular size cut-off. Changes were made twice every 3 hours after which they were dispensed into convenient volumes and stored in Eppendorf tubes at -20 C. DNA concentration was spectrophotometrically measured at 260/280 nm wavelengths; we standardized the amount of DNA that yielded maximum nicking in our experiments and established a working range of 2-6$\times$10$^{11}$ molecules in 300 $\mu\ell$ of sample volume, with the lower end of this range yielding the best percentage of relaxed species post laser exposure. The concentration of DNA plasmids was in the range 1.9-3.8$\times$10$^{11}$ cm$^{-3}$, corresponding to concentrations of 0.9-1.8 $\mu$g per 300 $\mu\ell$. We estimate that out of these molecules, 3$\times$10$^8$ are located within the interaction volume, constituting 0.03\% of plasmids.  Moreover, propagation of the laser beam through the cuvette containing DNA+water sets up thermal gradients that are strong enough to cause convective flow: a fresh set of DNA molecules constantly enters the laser-interaction zone in the course of a typical 2-minute exposure time. At the highest incident intensities, even bubbles are generated (as imaged in Fig. 1a that depicts the interaction geometry). We note that the amount of the energy deposited in a {\em single} filament, at a clamped intensity of 10$^{13}$ W cm$^{-2}$, is $\sim$1.5 $\mu$J. Post irradiation, electrophoresis enabled separation of DNA fragments by size. After separation, the gel was stained with a DNA binding fluorescent dye, ethidium bromide, to enable viewing by a BIORAD Gel Documentation system. Fragment size determination was accomplished with reference to commercially available DNA ladders containing linear fragments of known length. Since electrophoresis is used for assessment of DNA damage, the documented gels were used for measurement and analysis using standard gel-analysis software (ImageJ). 

Typical data for percentage change in supercoiled DNA upon irradiation with our laser pulses is shown in Fig. 1b along with a typical supercontinuum spectra that we measured (Fig. 1c) with a fiber-based spectrometer over the spectral range 200-870 nm. Asymmetric broadening towards the blue of the supercontinuum provides ready confirmation of plasma formation within the irradiated medium \cite{whitelight}. Is the dramatic increase that is observed in fraction of relaxed DNA due to photodamage? We note that the bluest part of the supercontinuum clamps at $\sim$400 nm; the absence of 266 nm radiation (where DNA absorbs most efficiently) rules out single-photon damage to DNA. Furthermore, on the basis of slopes obtained in log-log plots of percentage DNA damage as a function of laser intensity, we rule out 2- and 3-photon induced damage. The use of gel electrophoresis allows us to quantify the extent of DNA damage and relate it to parameters like irradiation time (akin to the radiation dose), DNA concentration, and laser energy. 

\begin{figure}
\includegraphics[width=8cm]{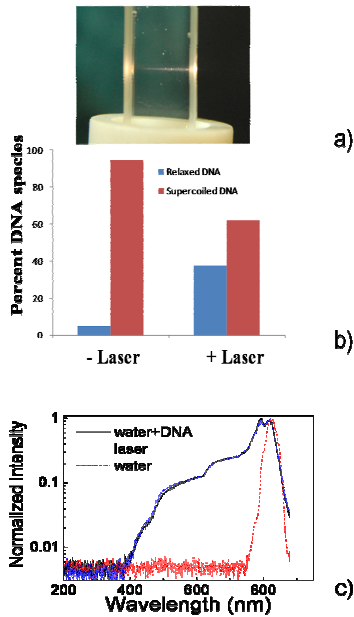}
\caption{(Color online) a) The laser-DNA interaction zone is imaged as a conical volume (base width $\sim$125 $\mu$m, length 1 cm) defined by multiple filamentation that occurs when an intense, focused laser beam propagates through a quartz cuvette containing DNA+water. At the highest incident laser intensities, bubbles are also formed, as seen. b) Percentage of supercoiled and relaxed DNA in normal conditions (- Laser) and after irradiation (+ Laser). c) Spectra of white light generated upon irradiation of water and water+DNA with 820 nm light of intensity 5 TW cm$^{-2}$. The narrower spectrum of the incident laser beam is also shown. Note the logarithmic scale.}
\end{figure}

Figure 2 shows data measured as the exposure time was varied over the range 10-120 seconds for fixed concentration of DNA and laser energy (130 $\mu$J). The gel electrophoresis data and the corresponding graphical representation show that the percentage of relaxed species increases to $\sim$15\% at 120 s exposure time. Using this exposure, we find that as DNA concentration was varied over the range 2-6 $\mu$g/$\mu\ell$, $\sim$15\% to 66 \% of the supercoiled DNA was converted to relaxed form. The yield of relaxed species in all plasmids varied from 10\% to 33\% as the laser intensity was increased from 1 to 4 TW cm$^{-2}$. 

\begin{figure}
\includegraphics[width=8cm]{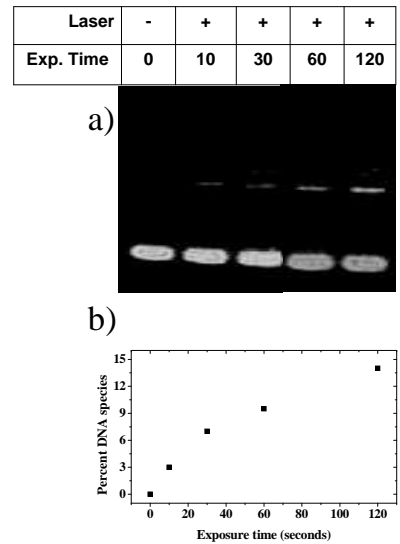}
\caption{a) Gel electrophoresis data for DNA plasmid pBR322 exposed for periods of times ranging from 0 to 120 seconds. The lower bright image denotes supercoiled DNA while the upper traces that become visible at 10 s and more prominent after 60 s denote relaxed DNA. b) Graphical quantification of the gel data. The DNA concentration was 2 $\mu$g/$\mu\ell$ and the laser intensity was 3 TW cm$^{-2}$.}
\end{figure}

We identify three possible processes that might set in as the intense laser pulse propagates through the DNA+water solution:
\begin{enumerate}
	\item The laser light (800 nm) can be absorbed by the DNA through three-photon absorption (direct process).
	\item Multiphoton ionization of water can generate free electrons \cite{myPRL} which become solvated electrons and react with proximate H$_2$O to form H and OH$^-$ anions. H-atoms thus formed can induce strand breaks; free electrons may also interact with DNA to break stands.
	\item The high optical field ionizes water molecules which yield, after efficient proton transfer, H$_3$O$^+$+OH radials. 
\end{enumerate}
 
Plasma formation in water that has been irradiated by intense laser light has been well studied (for a review, see \cite{vogel}) and the breakdown process has been modeled \cite{saachi} by treating water as an amorphous semiconductor with 6.5 eV band gap. Moreover, nonlinear absorption of liquid water not only involves ionization but also dissociation of the water molecules, leading to formation of reactive radicals. The quasi-free electrons that are produced gain further energy from the optical field via inverse bremsstrahlung and participate in further ionizing collisions. Rate equations for optical breakdown in water indicate that electron densities of 10$^{18}$-10$^{20}$ cm$^{-3}$ can be attained \cite{vogel,shen} and, we postulate, that it is these electrons that contribute to formation of temporary negative ions in water+DNA. The breakup of such negative ions results in strand breakages \cite{Li,Huels}. 

\begin{figure}
\includegraphics[width=8cm]{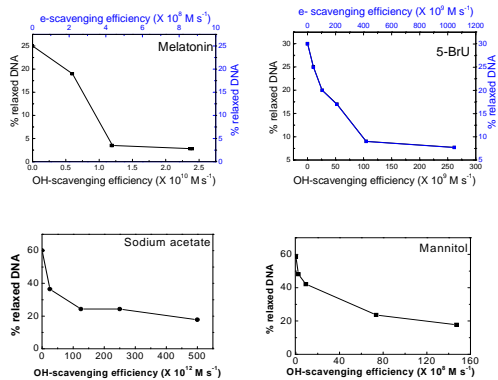}
\caption{Percentage of relaxed DNA as a function of e- and OH-scavenging efficiencies upon addition of (top two panels) melatonin and 5-bromouracil (5BrU) and (bottom two panels) OH-scavenging efficiences upon addition of sodium acetate and mannitol. The laser intensities were not the same for all panels. Conversion of measured scavenger concentrations into scavenging efficiences used rate coefficients given in \cite{EPAPS}.}
\end{figure}

To ascertain whether the relaxed form of DNA post-laser exposure in our experiments was, indeed, mediated by {\it in-situ} production of free electrons and radicals, we carried out experiments wherein various quenchers of electrons and radicals were added to the water-DNA sample (Fig. 3). We made measurements when water+DNA was doped with 5-bromouracil or 5-bromo-2,4(1H,3H)-pyrimidinedione), melatonin (N-acetyl-5-methoxytryptamine), which are both electron (primarily) as well as radical scavengers and sodium acetate and mannitol that scavenge mainly OH radicals \cite{PNAS}. These electron and radical scavengers were chosen with care to ensure that they do not chemically react with DNA to induce strand breakages. To place our doping levels in perspective, we note that in the case of melatonin, for instance, we typically used 2 mM concentration, which corresponds to 2.8$\times$10$^{17}$ molecules per cc out of which as many as 8.4$\times$10$^{13}$ are within the filamentation-determined interaction volume (Fig. 1a). Conversion of doping concentrations into scavenging efficiences relied on measured rate coefficients (see \cite{EPAPS} and references therein for values and associated errors). Our results (Fig. 3) present clearcut evidence that upon removal of electrons and/or radicals, the extent of DNA damage is significantly curtailed. We note that the percentage relaxation of DNA does not go to zero for the largest concentrations of each set of scavengers, implying that both electrons and radicals are responsible for strand breakages. Careful interpretation of data in the four panels offers indications that radical-induced damage is most likely the dominant strand-breakage mechanism. For instance, the amount of reduction in relaxed DNA species upon doping with mannitol or sodium acetate indicates that radicals may well account for about 80\% of the damage, the remaining 20\% being ascribable to electrons.

In summary, our experiments on laser-DNA interactions in the liquid phase have (i) demonstrated a qualitatively new approach to generate, {\it in-situ} electrons and radicals in an aqueous environment; (ii) both electrons and radicals interact with DNA plasmids kept in physiologically-relevant conditions so as to produce nicks in the plasmid DNA; and (iii) the number of nicks thus produced is measured to be directly proportional to the laser exposure. This study has implications that extend beyond DNA; the method we have succeeded in adopting here is likely to be applicable to studies of processes that are mediated by electrons and/or neutral radicals in many environments.

We thank Leon Sanche for suggesting the use of 5-bromo-uracil as an appropriate electron scavenger and we also gratefully acknowledge useful discussions with Arnaud Couairon on electron density estimates.

\end{document}